\documentclass[twocolumn,tighten]{aastex62}
\usepackage{amssymb, amsmath,framed}

\usepackage{bm}
\expandafter\ifx\csname package@font\endcsname\relax\else
 \expandafter\expandafter
 \expandafter\usepackage
 \expandafter\expandafter
 \expandafter{\csname package@font\endcsname}
\fi
\def\bq{\begin{equation}}
\def\eq{\end{equation}}
\def\bqy{\begin{eqnarray}}
\def\eqy{\end{eqnarray}}

\begin{document}
\title{\large{Dependence of Biological Activity on the Surface Water Fraction of Planets}}

\correspondingauthor{Manasvi Lingam}
\email{manasvi.lingam@cfa.harvard.edu}

\author{Manasvi Lingam}
\affiliation{Institute for Theory and Computation, Harvard University, Cambridge MA 02138, USA}
\affiliation{Harvard-Smithsonian Center for Astrophysics, Cambridge, MA 02138, USA}

\author{Abraham Loeb}
\affiliation{Institute for Theory and Computation, Harvard University, Cambridge MA 02138, USA}
\affiliation{Harvard-Smithsonian Center for Astrophysics, Cambridge, MA 02138, USA}

\begin{abstract}
One of the unique features associated with the Earth is that the fraction of its surface covered by land is comparable to that spanned by its oceans and other water bodies. Here, we investigate how extraterrestrial biospheres depend on the ratio of the surficial land and water fractions. We find that worlds that are overwhelmingly dominated by landmasses or oceans are likely to have sparse biospheres. Our analysis suggests that major evolutionary events such as the buildup of O$_2$ in the atmosphere and the emergence of technological intelligence might be relatively feasible only on a small subset of worlds with surface water fractions ranging approximately between $30\%$ and $90\%$. We also discuss how our predictions can be evaluated by future observations, and the implications for the prevalence of microbial and technological species in the Universe.\\
\end{abstract}

\section{Introduction} \label{SecIntro}
With the rapid advances in exoplanetary science over the past few years, it has become apparent that extrasolar worlds are very diverse in their physical properties \citep{WF15}. Statistical analyses of the mean density of exoplanets indicate that the majority of super-Earths above a certain radius have high fractions of volatiles \citep{Rog15,WL15,ChKi17,ZJS18,JM18}, although the exact magnitude of this cutoff is subject to some variability. To consider a more specific example, the recent discovery of seven Earth-sized planets transiting the ultracool dwarf star TRAPPIST-1 \citep{GT17} has led to theoretical modeling of the putative water contents of these planets \citep{GD18,UD18,DMG18}. Although the exact magnitude of the water inventories has varied from study to study, the water fraction (by weight) for some of these planets may reach a maximum of $40$-$50\%$ \citep{UHD18}. Hence, it has become apparent that many worlds will have surfaces with deep oceans and completely sans landmasses. 

In contrast, the marbled appearance of Earth's present-day surface, with its landmasses and oceans, has long been appreciated from the standpoint of aesthetics \citep{Sag94}. The Earth's distinctive topography is also characterized by an unusual coincidence: the fraction of the surface covered by landmasses ($29\%$) is comparable to the fraction encompassed by water bodies ($71\%$). Moreover, these two fractions were possibly comparable to one another since at least $3$ Gyr ago \citep{HCD17}; at the same time, it must be acknowledged that this issue is subject to several uncertainties \citep{AN12}. However, one should bear in mind that some of the other well-known ``coincidences'' associated with the Earth have been argued to be the outcome of observation selection effects \citep{Bos02}, and therefore are not merely chance events as they are subject to inherent anthropic bias.

One of the most famous analyses based on anthropic reasoning concerns the apparent coincidence in the timescales for the emergence of technological intelligence and the main-sequence lifetime of the Sun \citep{Cart83}. It was thus argued by \citet{Cart83} that the timescale for technological intelligence is much longer than the stellar lifetime in general, implying that the former is very rare in the Universe. Although Carter's argument has been critiqued on multiple fronts \citep{Liv99,CVD}, it represents a classic example of how anthropic reasoning may be employed. The second example that springs to mind is the fact that the tidal forces exerted by the Moon and the Sun are comparable to one another. \citet{Bal14} proposed that this feature enabled the rise of tetrapodomorphs during the Devonian period via the formation of pools due to tidal modulation, and that this coincidence in the magnitudes of tidal forces could have an anthropic bias.\footnote{In addition, planets in the TRAPPIST-1 system may also display some degree of tidal modulation \citep{MaL18}.} Other anthropic consequences associated with the presence of a large moon have been discussed in \citet{WaB00} and \citet{Wal06}.

In a similar spirit, the coincidence of the land and water (mostly prevalent as oceans) fractions at the Earth's surface was recently investigated by \citet{Sim17}. This study employed Bayesian reasoning to conclude that the observed coincidence can be traced to observation selection effects, and that most worlds are likely to have greater expanses of water on the surface compared to the Earth. In this paper, we will investigate how certain features of extraterrestrial biospheres, with microbial or technological life, may depend upon the surficial land ($f_\ell$) and water ($f_w$) fractions of habitable worlds.\footnote{Subsurface biospheres are excluded from our analysis, despite their unquestionable importance, as worlds with such biota are not expected to produce detectable biosignatures \citep{Col14}.} In the event that the surface comprises permanent icy regions, we will treat them as landmasses; to put it differently, ``water'' refers solely to liquid H$_2$O herein. Another parameter of interest that we shall utilize often is the ratio of these two fractions ($\delta_{w}$):
\begin{equation}\label{deldef}
    \delta_{w} = \frac{f_\ell}{f_w} = \frac{1-f_w}{f_w},
\end{equation}
where the last equality follows from $f_\ell = 1- f_w$. For the Earth, as remarked earlier, the water fraction equals $f_\oplus \approx 0.7$ while the land-water ratio is $\delta_\oplus \approx 0.4$. Thus, while $\delta_{w}$ theoretically ranges between $0$ and $\infty$, it is of order unity for the Earth. 

The paper is organized as follows. In Sec. \ref{SecExBio}, we examine the biological potential of extraterrestrial biospheres and discuss the consequences for the buildup of O$_2$ in the atmosphere. In Sec. \ref{SecMTELW}, we discuss how the emergence of technological intelligence may depend on the surficial land-water ratio of these worlds. We follow this up with a discussion of how common are worlds with surface landmasses and oceans in Sec. \ref{SecComm}. We conclude with a summary of our key results in Sec. \ref{SecCon}.

\section{The productivity of extraterrestrial biospheres}\label{SecExBio}
As noted in Sec. \ref{SecIntro}, $\delta_{w}$ can be very small or very large on other worlds. Consider the case where most of the surface is comprised of land, with the fraction of water being very low. The habitability of desert worlds low in surface water content has been extensively investigated \citep{AA11,LF13,ZS13}, and there are theoretical grounds for believing that these worlds are capable of maintaining a stable climate. However, it seems reasonable to suppose that many of these desert worlds would be characterized by low biomass densities. Thus, in the limit $\delta_{w} \gg 1$, we will hypothesize that the availability of water serves as the limiting factor for putative biospheres.

Now, consider the opposite scenario in which the planet's surface is almost entirely covered by oceans. These worlds are often referred to as ocean planets or waterworlds \citep{Kuc03,LS04}, and several studies indicate that they are potentially habitable \citep{KSR13,NSR17,KF18,RL18}. However, one should keep in mind the fact that ``life as we know it'' also necessitates certain bioessential elements such as phosphorus (P). On Earth, phosphorus is involved in biological functions in the form of phosphates \citep{West87,KSP13}, but the major phosphorus minerals (e.g. apatites) are characterized by their relatively low solubility in water \citep{SB13}. Hence, if one considers phosphorus as a bioessential element for extraterrestrial life, it is conceivable that the availability of dissolved P, typically in the form of phosphates, constitutes the limiting factor for the biological productivity of the oceans \citep{Tyrr99,SG06,Fil08} in the limit $\delta_{w} \ll 1$ \citep{WP13}. On Earth, some tentative evidence indicates that the emergence of animals during the late Proterozoic eon may have been linked to fundamental changes in the phosphorus cycle \citep{Kn17,RP17}. Moreover, the limited availability of P was probably responsible for suppressing ocean productivity during the Precambrian period \citep{BC02,KS17} with respect to the modern era, although it was still significant in absolute terms. 

There is an important distinction that merits further elaboration at this juncture. The limit $f_w \rightarrow 1$ does \emph{not} imply that the ocean planets will be devoid of life completely. This is because a finite supply of nutrients is still accessible from submarine weathering, thereby allowing the respective biogeochemical cycles to function. For example, during the Archean eon, characterized by a lower land fraction with respect to the present, the net primary productivity of Earth's oceans may have been a few percent of the current value \citep{Can05,SB13}; while this comes across as rather ``small'' by today's standards, it nevertheless resulted in a sizable biosphere. Moreover, recent studies indicate that waterworlds are potentially capable of sustaining biospheres over Gyr timescales \citep{KF18} and that they could possess favorable thermodynamic conditions (e.g. freeze-thaw cycles) for the origin of life \citep{RL18}. In contrast, if one considers the opposite limit ($f_w \rightarrow 0$), the absence of surface liquid water would ostensibly make it much more difficult for surficial ``life as we know it'' to exist. 

In other words, one extreme is dominated by the access to liquid water and the other by the availability of P. We will suppose that biospheres that lie in between these two extremes possess a biological potential that is governed by the superposition of these two factors. 

\subsection{The land fraction and biological productivity}\label{SSecLF}
In the regime of $f_w \ll 1$, we have argued that the biological productivity of terrestrial biospheres is regulated by the access to water. It is instructive to consider the major components of Earth's contemporary hydrological cycle \citep{Cha92,Bern12}. A close inspection of Figure 1 of \citet{OK06} reveals that the major source of precipitation onto land is due to the evaporation of oceans. The rate of evaporation $\dot{R}_E$ (in kg/yr) is given by
\begin{equation}
    \dot{R}_E \sim f_w \Phi_\mathcal{E} \left(4\pi R^2\right),
\end{equation}
where $\Phi_\mathcal{E}$ is the evaporation flux (in kg m$^{-2}$ yr$^{-1}$) and $R$ is the radius of the planet. In general, $\Phi_\mathcal{E}$ is a complex function of the environmental parameters such as the wind speed and the ambient temperature. It could, for instance, be calculated by using the Penman equation or some variants thereof \citep{PT72,Shut07}; an even simpler version was derived in \citet{Fey63}. However, for the purpose of our simplified analysis, we assume that all external parameters are held fixed, effectively rendering $\Phi_\mathcal{E}$ constant. 

The water that has been evaporated will be precipitated over land and oceans in proportion to their area. In other words, the amount of H$_2$O that will be precipitated over land is $\dot{R}_L = f_l \cdot \dot{R}_E$, with $f_l = 1 - f_w$ denoting the land fraction. However, it is evident that the amount of precipitation received will be non-uniform. Consider a toy model in which a fraction $f_h$ of the land receives precipitation, whereas a fraction $(1-f_h)$ is completely arid. Loosely speaking, we can interpret $(1-f_h) \equiv f_d$ as the fraction of the land covered by deserts, i.e. regions that do not obtain precipitation in the form of liquid water. We introduce the precipitation flux $\Phi_\mathcal{P}$ (in kg m$^{-2}$ yr$^{-1}$) in the non-arid regions; note that the corresponding flux in the deserts is treated as being negligible in comparison. In this case, the global precipitation rate $\dot{R}_P$ (in kg/yr) over the land becomes
\begin{equation}
    \dot{R}_P \sim f_h f_l \left(\Phi_\mathcal{P}\right) \left(4\pi R^2\right) + \left(1-f_h\right) f_l \cdot (0) \cdot  \left(4\pi R^2\right),
\end{equation}
where it must be reiterated that $f_h$ denotes the fraction of land that receives water influx, and is therefore habitable if all other factors are favorable. Let us further suppose that $\Phi_\mathcal{P}$ and $\Phi_\mathcal{E}$ are similar in magnitude ($\Phi_\mathcal{P} \sim \Phi_\mathcal{E}$). The total amount of precipitation on land per unit time ($\dot{R}_P$) ought to equal the fraction of total water evaporated that is deposited on land per unit time ($\dot{R}_L$). From the preceding relations, we find
\begin{equation}\label{fhfw}
f_h \approx f_w,
\end{equation}
implying that $f_d \approx 1 - f_w$. For the Earth, it is known that $f_d \approx 0.33$ and $1-f_w \approx 0.3$, implying that the relation $f_d \approx 1 - f_w$ is valid. Thus, despite the many idealizations involved, this model still yields a surprisingly accurate result for $f_h$ on Earth. This prescription can be easily generalized by adopting the ansatz $f_h = f_w^{1+ \alpha}$, with $\alpha = 0$ representing the fiducial value because of its relative accuracy \citep{Sim17}. Therefore, we shall work with (\ref{fhfw}) henceforth for the sake of simplicity.

Since deserts are typically distinguished by their low precipitation and much lower production rates of organic compounds per unit area compared to other habitats \citep{HS81}, it must be noted that the generation of biomass on land approximately encompasses a fraction $f_h \cdot f_\ell$ of the planetary surface area. A common metric for measuring the biological potential of a biosphere is the net primary productivity (NPP), which quantifies the net amount of carbon fixed via photosynthesis. On average, the NPP for deserts is roughly an order of magnitude lower than savannas, whose characteristic NPP is $\sim 0.7$ kg m$^{-2}$ yr$^{-1}$ \citep{FB98,JJ00}. Among deserts, the NPP spans nearly two orders of magnitude as one transitions from the most extreme deserts with NPP of $\sim 3 \times 10^{-3}$ kg m$^{-2}$ yr$^{-1}$ to semi-deserts that have NPP of $\sim 0.2$ kg m$^{-2}$ yr$^{-1}$ \citep{Stil01,Dok13}.

In order to determine the NPP of the land and oceans, a detailed knowledge of the metabolic pathways and densities of putative extraterrestrial organisms is necessary. Clearly, we do not possess any knowledge of these aspects, owing to which we shall assume that the NPP per unit area (in kg m$^{-2}$ yr$^{-1}$) is commensurate with that of the Earth. Thus, we find that the NPP on land, denoted by $\mathcal{B}_\ell$ (in kg/yr), is given by
\begin{equation}\label{NPPL}
 \mathcal{B}_\ell \sim 5.6 \times 10^{13}\,\mathrm{kg/yr}\,\left(\frac{f_w}{f_\oplus}\right)\left(\frac{1 - f_w}{1 - f_\oplus}\right)\left(\frac{R}{R_\oplus}\right)^2,   
\end{equation}
where the normalization factor on the right-hand-side corresponds to the terrestrial NPP on Earth \citep{FB98} and $f_\oplus$ has been defined in Sec. \ref{SecIntro}. In the same spirit, if we assume that the mean biological turnover time of extraterrestrial producers (autotrophs) is similar to the Earth, the total biomass of these lifeforms on land (${M}_\ell$) can be expressed as
\begin{equation}\label{BioML}
{M}_\ell \sim 4.5 \times 10^{14}\,\mathrm{kg}\,\left(\frac{f_w}{f_\oplus}\right)\left(\frac{1 - f_w}{1 - f_\oplus}\right)\left(\frac{R}{R_\oplus}\right)^2,     
\end{equation}
where the normalization factor corresponds to the biomass of terrestrial primary producers on Earth \citep{BPM18}. An inspection of (\ref{NPPL}) and (\ref{BioML}) reveals that these quantities become infinitesimally small in the limits $f_w \rightarrow 0$ and $f_w \rightarrow 1$, when the worlds either lack surface water altogether or possess only oceans, in which case land-based ecosystems are automatically absent.

\subsection{The water fraction and biological productivity}\label{SSecWaP}
We have proposed that the biological productivity of ocean-dominated worlds may be limited by the availability of P. In this scenario, the NPP of the oceans ($\mathcal{B}_w$), is expressible as
\begin{equation}\label{Bwdef}
  \mathcal{B}_w \propto \phi_P f_w R^2,  
\end{equation}
where $\phi_P$ is the average steady-state concentration of dissolved P, while the factor $f_w R^2$ is proportional to the area of the ocean. There is a subtle caveat associated with the above equation - it is valid only when the concentration of P lies sufficiently below its saturation value. In other words, $\phi_P$ must be replaced by $\phi_P/\left(\mathcal{K}_P + \phi_P\right)$ in the general case, where $\mathcal{K}_P$ is the Monod constant \citep{SG06}. In order to determine $\phi_P$, we follow the approach outlined in \citet{LL18}, where $\phi_P$ is computed from
\begin{equation}\label{SSConP}
\phi_P = \frac{\sum \mathcal{S}_P}{M_{oc} \sum \mathcal{L}_P},   
\end{equation}
where $M_{oc}$ is the mass of the ocean (in kg), $\mathcal{S}_P$ denotes a given source of P (in mol/yr), and $\mathcal{L}_P$ represents a particular sink of P (in yr$^{-1}$). There are a multitude of sources and sinks of P, and many of them are poorly constrained, owing to which we shall focus only on the primary mechanisms. 

A significant fraction of the dissolved P input into the oceans is due to continental weathering, and is transported via rivers. Based on Sec. \ref{SSecLF}, only a fraction $f_h$ of the land actually receives precipitation and it seems reasonable to assume that rivers flow only through this region. Hence, the riverine source of P is given by
\begin{equation} \label{SPriv}
\mathcal{S}_\mathrm{P} \sim \,3 \times 10^{10}\,\mathrm{mol/yr}\,\left(\frac{f_w}{f_\oplus}\right)\left(\frac{1 - f_w}{1 - f_\oplus}\right)\left(\frac{R}{R_\oplus}\right)^2,
\end{equation}
where the normalization factor has been adopted from \citet{Wall10}. Note that the above expression vanishes in the limiting cases when: (a) oceans are entirely absent (zero precipitation and river runoff) and (b) land does not exist (zero continental weathering). The factor $f_w \left(1-f_w\right)$ is present in the above equation because it encapsulates the fraction of non-arid land, namely $f_h \cdot f_l$. This can be simplified further to yield $f_w \left(1-f_w\right)$ after using (\ref{fhfw}) and the definition of $f_l$.

The next major source of P to the oceans is from the deposition of atmospheric dust, aerosols and volcanic ash. Clearly, this source depends on the total fraction of available land ($f_\ell$), and the resultant products are distributed over land and oceans in proportion to their respective areas; recall that we are interested in the fraction deposited in the oceans ($f_w$). Thus, \emph{in toto}, the magnitude of this source depends on the quantity $f_w \cdot f_l \equiv f_w\left(1-f_w\right)$. This ansatz vanishes in the limits $f_w \rightarrow 0$ and $f_w \rightarrow 1$ along expected lines because having only oceans or only land would negate this source. The atmospheric source of P becomes
\begin{equation} \label{SPatm}
\mathcal{S}_\mathrm{P} \sim \,1 \times 10^{10}\,\mathrm{mol/yr}\,\left(\frac{f_w}{f_\oplus}\right)\left(\frac{1 - f_w}{1 - f_\oplus}\right)\left(\frac{R}{R_\oplus}\right)^2,
\end{equation}
where the normalization factor represents the amount of soluble reactive P contributed by atmospheric sources on Earth \citep{BN00}. 

The third source that we consider is the submarine weathering of the ocean floor by seawater. This is actually a minor source on our planet owing to the fact that the weathering rate per unit area, leading to the generation of dissolved P, has an inverse exponential dependence on the pH \citep{AHF13}. In other words, given that the pH of seawater is around $2.4$ units higher than rainwater, the magnitude of this source is $\sim 100$ times lower than the riverine P influx, yielding
\begin{equation} \label{SPsub}
\mathcal{S}_\mathrm{P} \sim \,1.3 \times 10^{8}\,\mathrm{mol/yr}\,\left(\frac{f_w}{f_\oplus}\right)\left(\frac{R}{R_\oplus}\right)^2,
\end{equation}
with additional details concerning the derivation described in \citet{LL18}. The essential point to be noted here is that (\ref{SPsub}) vanishes when $f_w \rightarrow 0$, but it does not vanish when land is absent, i.e. in the limit $f_w \rightarrow 1$, because it does not depend on the presence of continental landmasses. In this study, we are primarily interested in worlds where water-rock interactions near the seafloor exist. This leads to the presence of water-rock interactions at the ocean floor, thereby also opening up the possibility that life arose in hydrothermal vents \citep{BH85,RBB14}. In the event that high-pressure ices prevent these interactions \citep{NH16}, the oceans are expected to become acidic, i.e. with a pH of $2$-$4$ \citep{LS18}.

There are three other sources of P that we will ignore for the following reasons. The first is volcanic activity, which we do not consider because it is very localized and unlikely to be a major player on a global scale \citep{BN00}. The second is the exogenous delivery of P in the form of schreibersite via meteorites. While this source was fairly important on early Earth, it is expected to become unimportant once the flux of impactors sharply declines at later epochs. Even if we restrict ourselves to the early Earth, the upper bound on soluble reactive P through this channel has been estimated to be $\sim 10^8$ kg/yr \citep{PH17}, which translates to $\sim 3.3 \times 10^9$ mol/yr and is therefore a few times smaller than (\ref{SPriv}) and (\ref{SPatm}). The last source entails weathering due to glaciers, but this process has not been studied in much detail. The magnitude of this source is, at most, comparable to the riverine and atmospheric sources \citep{Wall10}, indicating that our results are not likely to be much altered by including it. 

Next, we turn our attention to the sinks of P. Broadly speaking, they can be divided into two major categories: the burial of marine sediments and precipitation at hydrothermal vents \citep{PM07}; the magnitude of the former sink is about $3$ times higher than the latter on Earth \citep{WMM03}. The key point is that the sinks exhibit the following scaling \citep{LL18}:
\begin{equation}
    \sum \mathcal{L}_P \propto \left(\frac{M}{M_{oc}}\right),
\end{equation}
where $M$ represents the total mass of the world. The characteristic depletion timescale ($\tau_P \equiv 1/\sum \mathcal{L}_P$) vanishes in the limit $M_{oc} \rightarrow 0$, as there would be no oceanic P for the sinks to eliminate in this scenario. The constant of proportionality in this expression is connected to the mean residence time associated with the sinks of P on Earth, but this quantity is subject to some variability \citep{BN00}. By utilizing the preceding information concerning the sources and sinks in conjunction with (\ref{SSConP}), we obtain
\begin{equation}\label{SSPfin}
    \frac{\phi_P}{\phi_\oplus} \sim \left(\frac{f_w}{f_\oplus}\right)\left[\left(\frac{1 - f_w}{1 - f_\oplus}\right) + 3.3 \times 10^{-3} \right]\left(\frac{R}{R_\oplus}\right)^{-1.7},
\end{equation}
where $\phi_\oplus \approx 2$ $\mu$M represents the average concentration of dissolved P in Earth's oceans \citep{KS17}. This quantity follows from the fact that the total inventory of dissolved P is approximately $3 \times 10^{15}$ mol \citep{PM07}, while the mass of Earth's oceans is $1.4 \times 10^{21}$ kg. In order to arrive at (\ref{SSPfin}), we have also invoked the mass-radius scaling $M \propto R^{3.7}$ \citep{ZSJ16} for rocky planets.

Substituting (\ref{SSPfin}) in (\ref{Bwdef}) yields the NPP of the oceans:
\begin{eqnarray}\label{NPPO}
 \mathcal{B}_w &\sim& 4.9 \times 10^{13}\,\mathrm{kg/yr}\, \left[\left(\frac{1 - f_w}{1 - f_\oplus}\right) + 3.3 \times 10^{-3} \right] \nonumber \\
 && \quad \times \left(\frac{f_w}{f_\oplus}\right)^2\left(\frac{R}{R_\oplus}\right)^{0.3},
\end{eqnarray}
where the normalization corresponds to the NPP of Earth's oceans \citep{FB98}. The above equation implies that $\mathcal{B}_w \rightarrow 0$ when $f_w \rightarrow 0$ as expected. However, in the limit $f_w \rightarrow 1$, we see that $\mathcal{B}_w$ remains finite. This is because the inventory of dissolved P is non-zero, albeit very small, even in the absence of landmasses. We can proceed further by assuming that the characteristic biological turnover time in extraterrestrial oceans is comparable to the Earth to determine the total biomass of producers in the oceans ($M_w$), thereby obtaining
\begin{equation}\label{BioMO}
 M_w \sim 10^{12}\,\mathrm{kg}\,\left(\frac{f_w}{f_\oplus}\right)^2 \left[\left(\frac{1 - f_w}{1 - f_\oplus}\right) + 3.3 \times 10^{-3} \right]\left(\frac{R}{R_\oplus}\right)^{0.3},
\end{equation}
and the normalization constant signifies the biomass of primary producers in Earth's oceans \citep{BPM18}. If we compare (\ref{BioMO}) with (\ref{BioML}), it is seen that the latter is about two orders of magnitude higher than the former for the Earth.

\subsection{Total net primary productivity and biomass}\label{SSecNPPB}

\begin{figure*}
$$
\begin{array}{cc}
  \includegraphics[width=6.8cm]{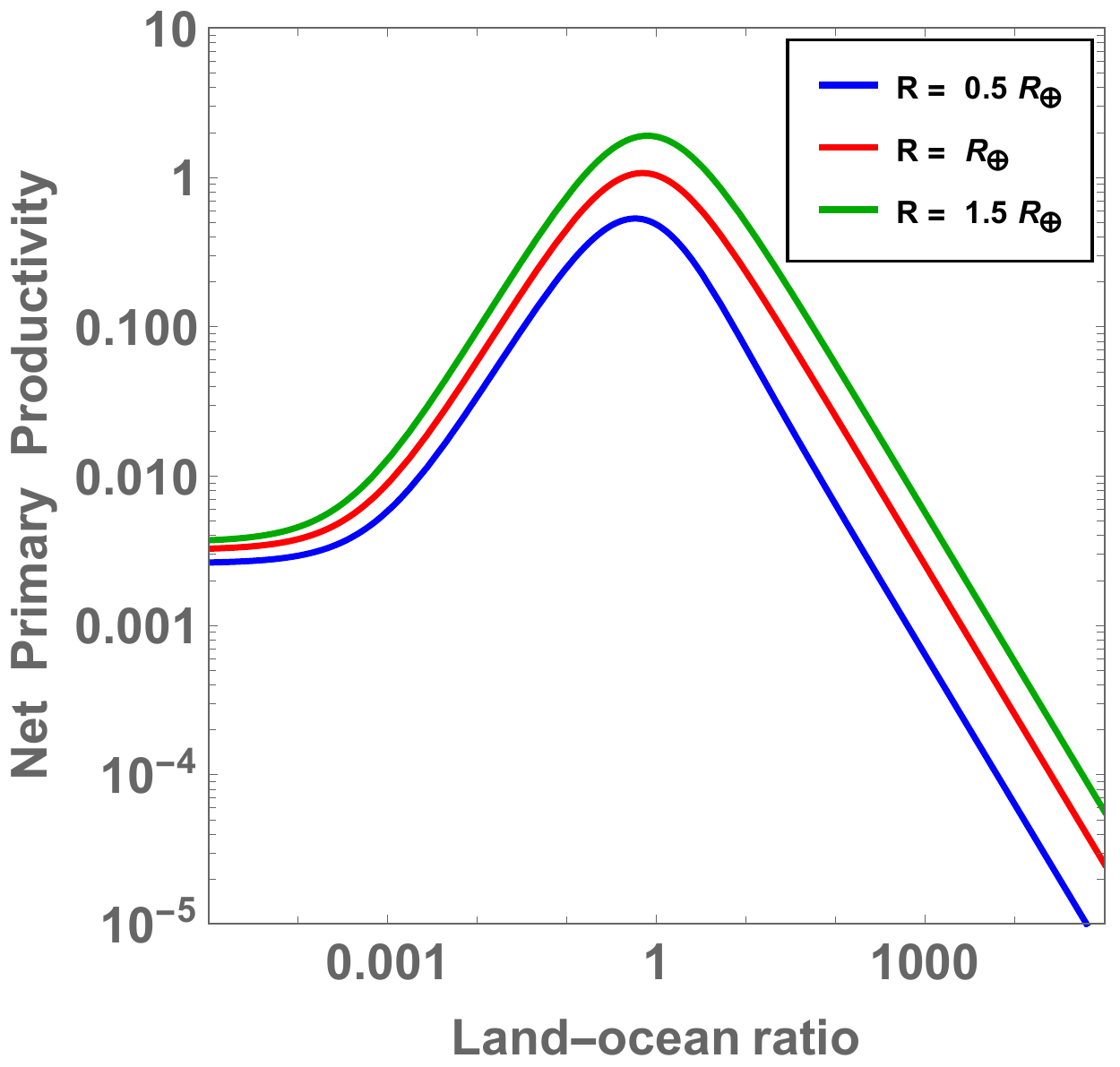} &  \includegraphics[width=6.8cm]{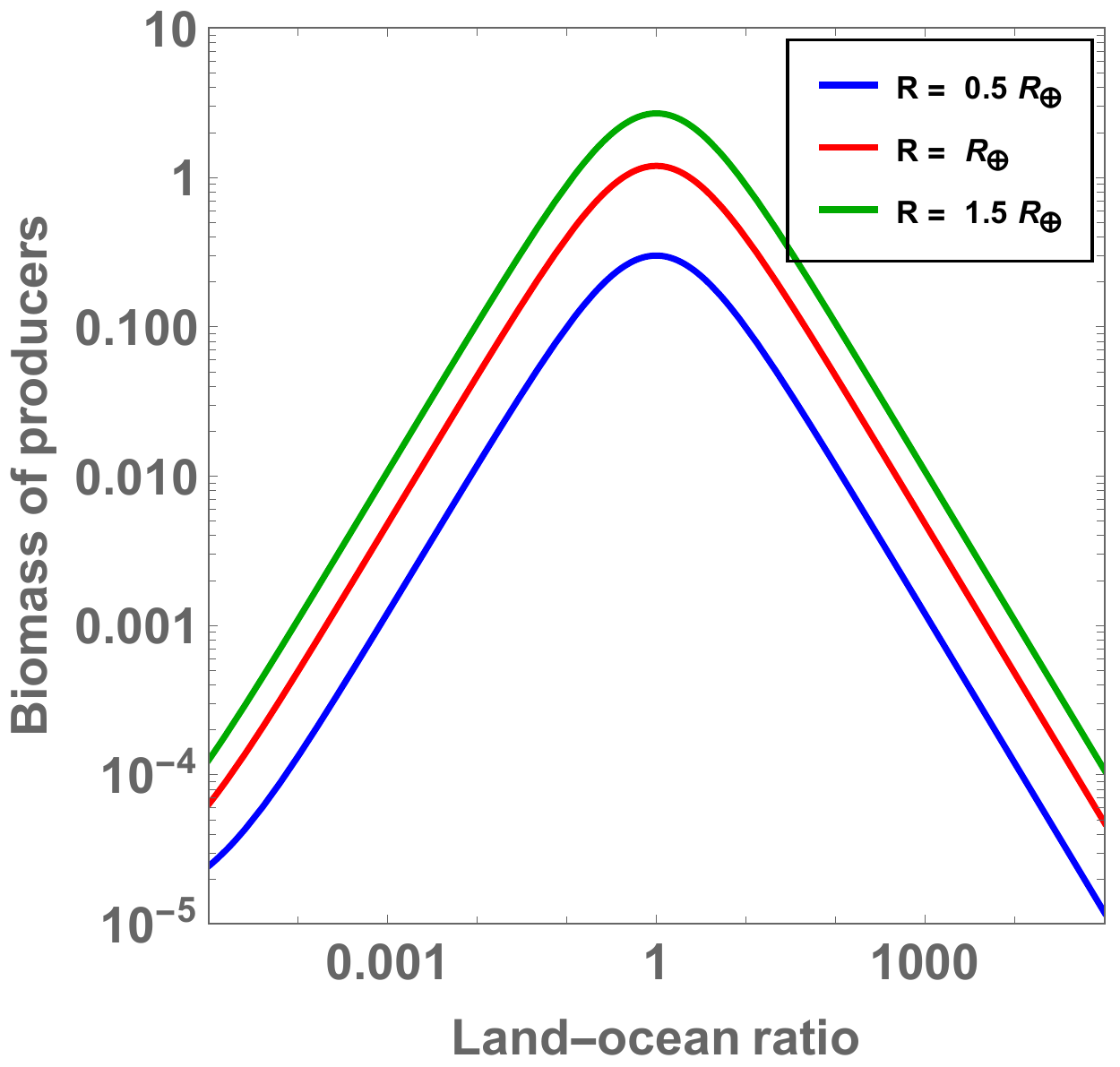}\\
\end{array}
$$
\caption{Left panel: The normalized Net Primary Productivity (NPP) relative to the Earth as a function of the land-ocean ratio. Right panel: The total biomass of primary producers (mostly photosynthetic organisms) normalized by its value for the Earth. In both panels, the horizontal axis denotes the ratio of the planet's surface covered by landmasses to that covered by oceans. The blue, red and green curves correspond to worlds with radii of $0.5\,R_\oplus$, $R_\oplus$, and $1.5\,R_\oplus$ respectively.}
\label{FigLWBio}
\end{figure*}

Based on the previous calculations, we can determine the total net primary productivity $\mathcal{B}_t = \mathcal{B}_\ell + \mathcal{B}_w$ and the total biomass of producers $M_t = M_\ell + M_w$. It is, however, more instructive to determine their values normalized to that of the Earth. 

Hence, we introduce the following ratios:
\begin{eqnarray}\label{DelBdef}
 \Delta_\mathcal{B} &=& \frac{\mathcal{B}_t}{\mathcal{B}_{t,\oplus}} \sim  f_w \left(\frac{R}{R_\oplus}\right)^2 \Bigg[2.52\left(1-f_w\right) \\
 && \quad + 3.2 f_w \left[\left(1-f_w\right) + 10^{-3} \right]\left(\frac{R}{R_\oplus}\right)^{-1.7} \Bigg] \nonumber,  
\end{eqnarray}
\begin{eqnarray}\label{DelMdef}
\Delta_M &=& \frac{M_t}{M_{t,\oplus}} \sim f_w \left(\frac{R}{R_\oplus}\right)^2 \Bigg[4.76\left(1-f_w\right) \\
&& \quad + 1.51 \times 10^{-2}f_w \left[\left(1-f_w\right) + 10^{-3} \right]\left(\frac{R}{R_\oplus}\right)^{-1.7} \Bigg]. \nonumber 
\end{eqnarray}
From these expressions, it is straightforward to estimate the $f_w$'s at which the maximum values of the $\Delta$'s are attained. Setting $d \Delta_\mathcal{B}/d f_w = 0$ yields
\begin{eqnarray}
f_\mathcal{B} &\approx& \frac{1}{3}\Bigg[1 - 0.79 \left(\frac{R}{R_\oplus}\right)^{1.7} \\
&&\,\, +\, \sqrt{1 + 0.79 \left(\frac{R}{R_\oplus}\right)^{1.7} + 0.62\left(\frac{R}{R_\oplus}\right)^{3.4}} \Bigg], \nonumber
\end{eqnarray}
where $f_\mathcal{B}$ is the water fraction at which the maximum value of $\Delta_\mathcal{B}$ is reached. Similarly, we can compute $f_M$ from $d \Delta_M/d f_w = 0$, thereby ending up with
\begin{eqnarray}
f_M &\approx& \frac{1}{3}\Bigg[1 - 315 \left(\frac{R}{R_\oplus}\right)^{1.7} \\
&&\,\,+\, \sqrt{1 + 315 \left(\frac{R}{R_\oplus}\right)^{1.7}
+ 99173\left(\frac{R}{R_\oplus}\right)^{3.4}} \Bigg]. \nonumber
\end{eqnarray}
For the case of $R = R_\oplus$, we find $f_\mathcal{B} \approx 0.59$ and $f_M \approx 0.5$. The corresponding $\Delta$'s are $\Delta_\mathcal{B} \approx 1.07$ and $\Delta_M \approx 1.19$. Thus, within the context of this simple model, we see that the net primary productivity and biomass of Earth's biosphere are very close to the peak values. From (\ref{deldef}), we see that $\delta_w\left(f_\mathcal{B}\right) \approx 0.7$ and $\delta_w\left(f_M\right) \approx 1$, indicating that the land-water fraction is close to unity in both cases.

We recall that the variable $\delta_w$ was introduced earlier because it can theoretically span many orders of magnitude. It is therefore instructive to plot (\ref{DelBdef}) and (\ref{DelMdef}) as a function of $\delta_w$. The results have been depicted in Fig. \ref{FigLWBio} for three different values of $R/R_\oplus$. For the case with $R = R_\oplus$, we find that $\Delta_\mathcal{B} > 0.1$ occurs when $\delta_w \in \left(0.02, 20\right)$, implying that $\delta_w$ spans three orders of magnitude. If we repeat the same analysis for $\Delta_{M} > 0.1$, we find that the corresponding domain is very similar, i.e. we obtain $\delta_w \in \left(0.02, 50\right)$. In the left panel of Fig. \ref{FigLWBio}, we see that the curves flatten as $\delta_w \rightarrow 0$. This behavior arises from the fact that (\ref{DelBdef}) does not vanish in the limit $f_w \rightarrow 1$ (equivalent to $\delta_w \rightarrow 0$) owing to the supply of dissolved phosphorus through the weathering of the ocean floor. This flattening also applies to (\ref{DelMdef}), but this trend is not manifested in the right panel of Fig. \ref{FigLWBio} because $\delta_w$ does not reach sufficiently small values.

\subsection{Consequences for the buildup of oxygen}
At this stage, recall that the total NPP ($\mathcal{B}_t$) represents the rate of organic carbon fixation primarily via photosynthesis. An important point to keep in mind is that photosynthesis should not automatically be equated with \emph{oxygenic} photosynthesis, as several anoxygenic photosynthetic pathways exist that involve electron donors like Fe$^{2+}$ and H$_2$S \citep{XB02,HB11,KNo17}. Nonetheless, the evolution of oxygenic photosynthesis led to a dramatic change in the evolutionary landscape \citep{deDu,Knoll15,FHJ16,Jud17} and most of the NPP on Earth is currently dominated by oxygenic photosynthesis. Hence, we will operate under the premise that oxygenic photosynthesis represents the dominant mechanism for the synthesis of organic compounds on other worlds. 

On Earth, the amount of organic compounds (in the form of CH$_2$O) produced per year is about $3.5 \times 10^{15}$ mol/yr, derived from converting $\mathcal{B}_{t,\oplus}$ (in kg/yr) into mol/yr. It has been estimated that a small fraction of the total organic carbon is buried, and provides a net source of O$_2$. The upper bound on the burial efficiency is approximately $2.9 \times 10^{-3}$ \citep{Holl02}, and multiplying this with the production rate of organic matter yields the amount of O$_2$ produced. Thus, the net production rate of O$_2$ on a given world is
\begin{equation}\label{SO2bur}
    \mathcal{S}_{O_2} \sim 10^{13}\,\mathrm{mol/yr}\,\Delta_\mathcal{B},
\end{equation}
where $\Delta_\mathcal{B}$ has been defined in (\ref{DelBdef}). Another major source of O$_2$ is the burial of pyrite (FeS$_2$) as it would otherwise react with oxygen and lower its concentration. There are a couple of major sinks, but only one of them is prominent in a world where O$_2$ has not built up to sufficiently high levels, namely the consumption of O$_2$ due to rapid reactions with reducing gases produced from volcanism and submarine weathering of the ocean floor \citep{CK17}. Assuming that the flux of reducing gases is constant (units of mol m$^{-2}$ yr$^{-1}$), the depletion rate of O$_2$ can be written as
\begin{equation}\label{LO2red}
   \mathcal{L}_{O_2} \sim 5.7 \times 10^{12}\,\mathrm{mol/yr}\,\left(\frac{R}{R_\oplus}\right)^2,
\end{equation}
where the proportionality constant for the Earth has been adopted from \citet{CK17}. Although we have used the notation $\mathcal{L}$ in the above equation, it is vital to recognize that it has different units compared to $\mathcal{L}_P$ in Sec. \ref{SSecWaP}. It was recently proposed in \citet{LCP18} that a potentially sufficient condition for ensuring the buildup of oxygen is that the source term must exceed the depletion term, i.e. we require (\ref{SO2bur}) to exceed (\ref{LO2red}). After some simplification, we obtain the condition
\begin{eqnarray}\label{O2Ineq}
 \mathcal{G}\left(f_w,R\right)  > 0.57,
\end{eqnarray}
where the function $\mathcal{G}\left(f_w,R\right)$ is defined as
\begin{equation}
 \mathcal{G} \equiv 2.52f_w\left(1-f_w\right) + 3.2 f_w^2 \left[\left(1-f_w\right) + 10^{-3} \right]\left(\frac{R}{R_\oplus}\right)^{-1.7}.
\end{equation}

\begin{figure}
\includegraphics[width=7.5cm]{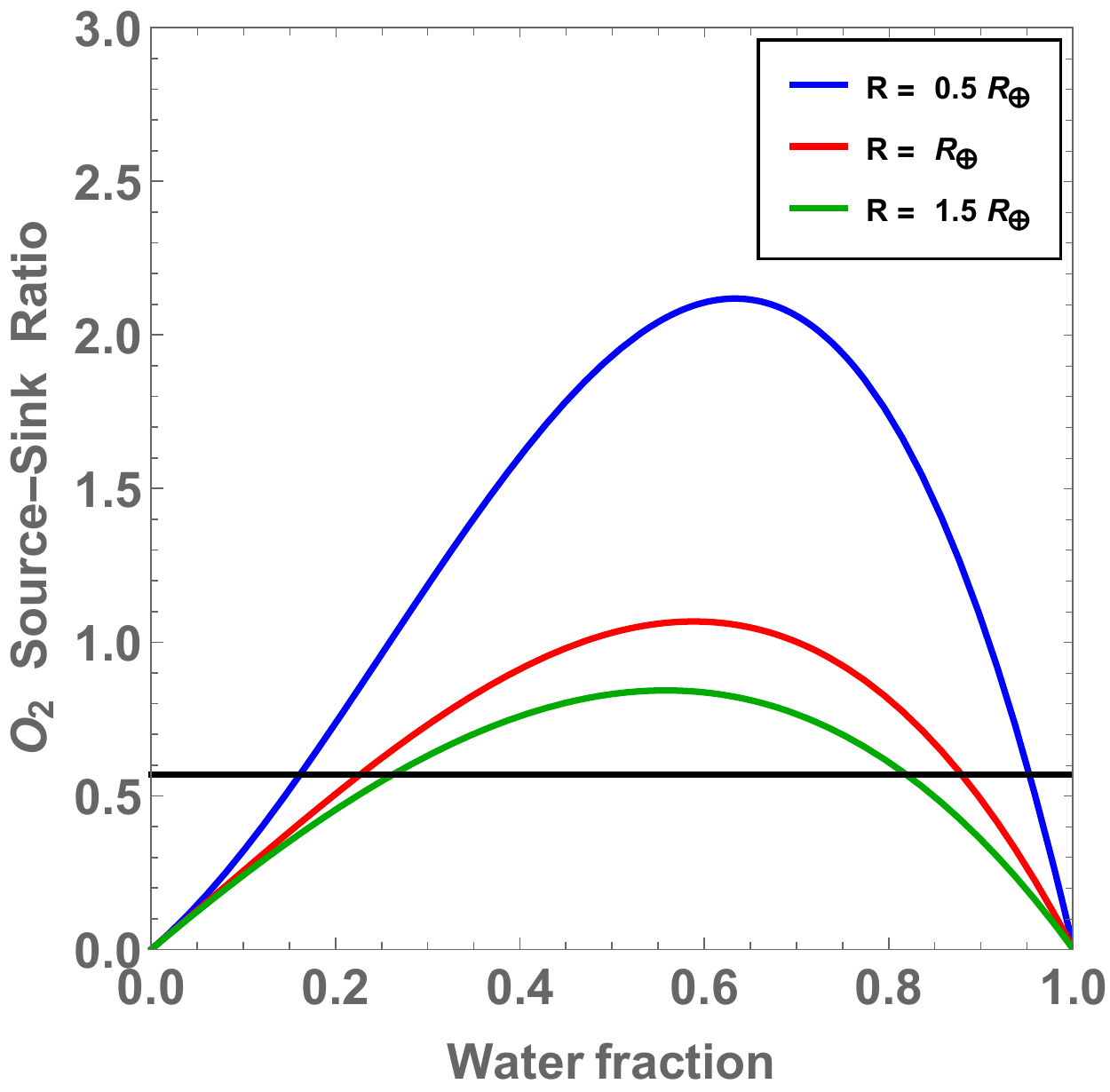} \\
\caption{The left-hand-side of (\ref{O2Ineq}), which represents the ratio of oxygen sources to sinks, as a function of the water fraction $f_w$ for different choices of the planetary radius $R$. When the curves cross the horizontal black line, the buildup of O$_2$ in the atmosphere could take place. The blue, red and green curves correspond to worlds with radii of $0.5\,R_\oplus$, $R_\oplus$, and $1.5\,R_\oplus$ respectively.}
\label{FigSSRat}
\end{figure}

The parameter space where this inequality is fulfilled can be found by inspecting Fig. \ref{FigSSRat}. The above criterion is satisfied within the regions where the curves cross above the black line. For $R = R_\oplus$, we find that (\ref{O2Ineq}) is valid only when $f_w \in \left(0.23,0.88\right)$ and the corresponding interval for $\delta_w$ becomes $\delta_w \in \left(0.14,3.35\right)$. Hence, it is apparent that the allowed values of $\delta_w$ fall within a fairly narrow range spanning roughly one order of magnitude. Another interesting result from Fig. \ref{FigSSRat} is that this range increases for smaller worlds, i.e. when $R$ is lowered.

We reiterate that our analysis is not meant to be exhaustive, as this does not include relatively minor sources and sinks on Earth that may turn out to be important on other worlds. Furthermore, our analysis assumes that extraterrestrial biogeochemistry is akin to that of the Earth, and it also neglects the fact that biogeochemical cycles are dynamical in nature. 

\section{Major transitions in evolution and the land-water ratio}\label{SecMTELW}

In order for technological intelligence to emerge on Earth, several theoretical frameworks indicate that a small number ($\lesssim 10$) of critical evolutionary steps were necessary \citep{JMS95,Koo7,La09,CS11,Sza15}. It is often posited that these transitions must occur independently, and in sequential order characterized by increasing biological complexity. The number and nature of these critical steps has been subject to debate, but several recent studies favor five to six transitions in total \citep{Cart08,Wat08,ML10,LiLo18}. 

There are two important caveats worth highlighting here. First, models reliant on critical steps are not the only viable explanation for the emergence of complex life (and technologically advanced life) on Earth and elsewhere. One possible alternative is that many paths may have led to each of these major evolutionary transitions, thereby implying that these events have a relatively high likelihood of occurrence, given enough time and the right environmental conditions \citep{SB17}. Second, there is no guarantee that the same critical steps would be necessary for the emergence of technological intelligence on other worlds, unless one subscribes to a high degree of evolutionary convergence \citep{Mor03}. Despite these issues, an inherent advantage associated with utilizing critical steps models of the major evolutionary transitions is that it they offer a useful approach for understanding the likelihood of technological intelligence on other worlds. 

Let us consider a particular evolutionary transition, namely, the origin of life (abiogenesis). It is common to model abiogenesis as the consequence of a very large number of independent, random ``trials''. A vital point worth appreciating here is that most studies are concerned with the temporal element, i.e. the timescale for abiogenesis \citep{Cart83,LD02,ST12}. In actuality, however, the spatial component is equally important, especially if one supposes that life originated in a local environment \citep{Deam97}. We know neither the characteristic area of this putative ``microenvironment'' nor its exact nature, but we may conjecture that the former equals $\mathcal{A}_\ell$ (land) or $\mathcal{A}_w$ (ocean). In case the microenvironments are situated in the ocean, the number of microenvironments ($\mathcal{N}_m$) is given by $\mathcal{N}_m \propto f_w R^2/\mathcal{A}_w$. In the same spirit, if the microenvironments are land-based, we have $\mathcal{N}_m \propto f_w\left(1 - f_w\right) R^2/\mathcal{A}_\ell$. The extra factor of $f_w$ is manifested because we are interested in the fraction of land that is ``habitable'', and we have also invoked $f_h \approx f_w$. In both cases, if we presume that the corresponding areas of these microenvironments are similar for Earth and other worlds, we have $\mathcal{N}_m \propto f_w R^2$ for oceans and $\mathcal{N}_m \propto f_w\left(1 - f_w\right) R^2$ for land. 

\begin{figure*}
$$
\begin{array}{cc}
  \includegraphics[width=6.8cm]{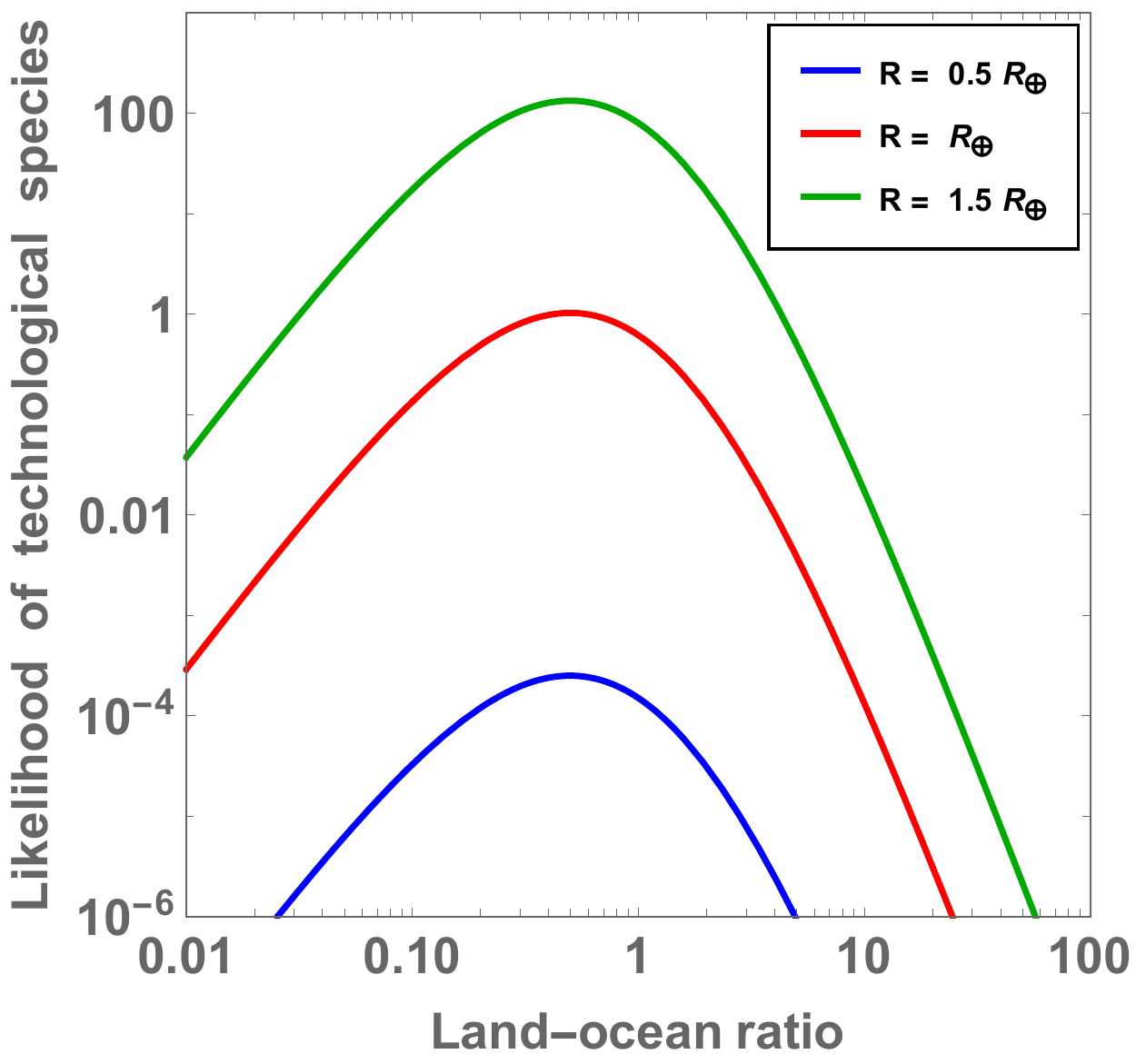} &  \includegraphics[width=6.8cm]{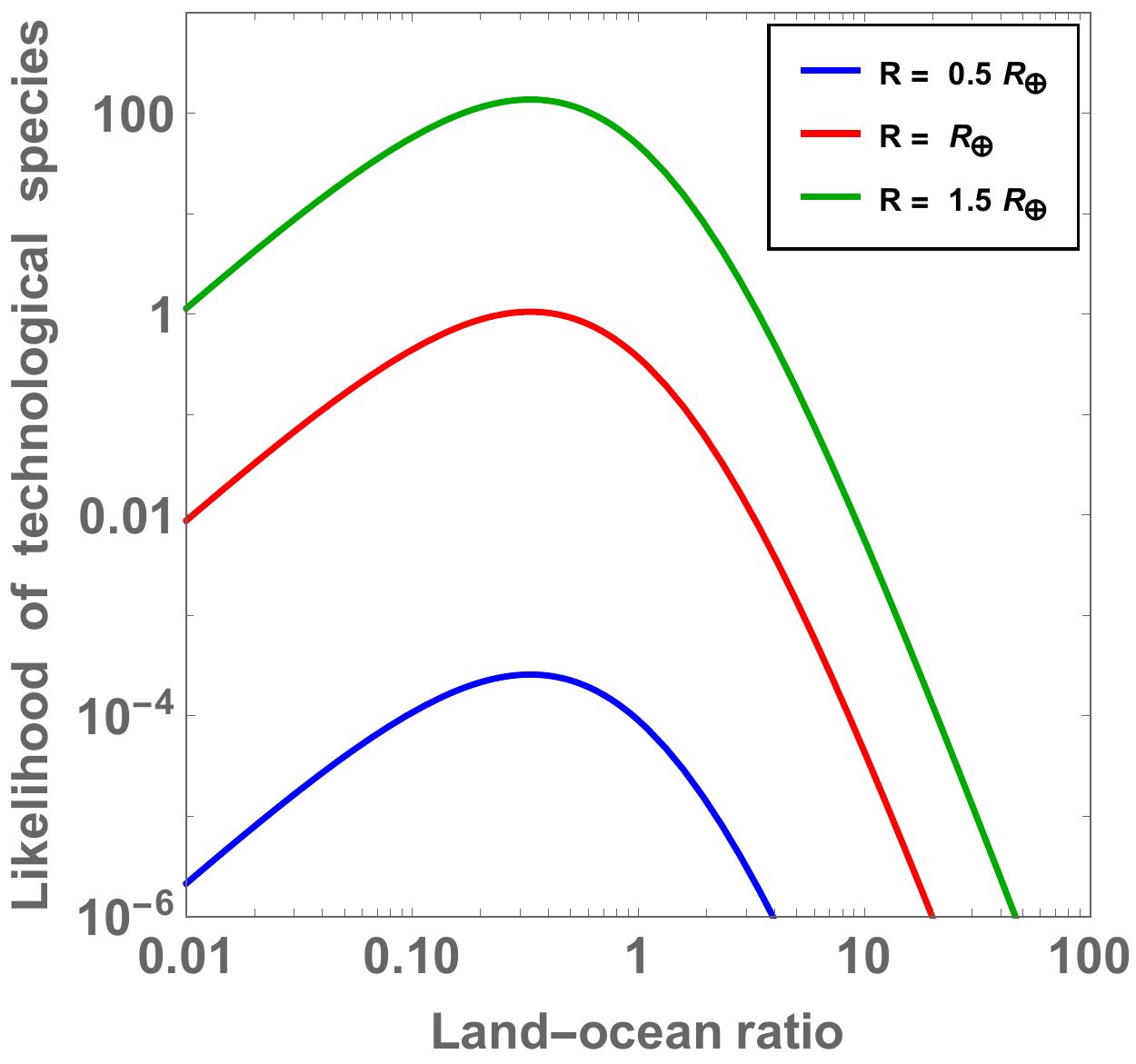}\\
\end{array}
$$
\caption{Left panel: Overall likelihood of technological intelligence relative to the Earth as a function of the land-ocean ratio, assuming that life originated in a terrestrial environment. Right panel: The same quantity assuming that life originated in an aquatic environment. In both panels, the $x$-axis denotes the ratio of the planet's surface covered by landmasses to that covered by oceans. The blue, red and green curves correspond to worlds with radii of $0.5\,R_\oplus$, $R_\oplus$, and $1.5\,R_\oplus$ respectively.}
\label{FigTLE}
\end{figure*}

Although we have focused on abiogenesis for the sake of illustrating our point, a similar line of reasoning could also apply to the other major evolutionary transitions. To complete our picture, we make use of $\mathcal{P}_j \propto \mathcal{N}_j$, where $\mathcal{P}_j$ is the likelihood of occurrence of the $j$-th critical step. This relation essentially implies that the probability of success will be proportional to the maximum number of spatial ``trials'' allowed, i.e. the number of accessible microenvironments \citep{deDu}. The literature comprises numerous critical step models, which are subject to variations in both the number and choice of the major evolutionary breakthroughs. We will adopt the ``megatrajectories'' framework \citep{KB00} that was recently reviewed in \citet{LiLo18} and argued to be a fairly credible critical steps model on theoretical grounds. There are six megatrajectories, which are described below.\\

\noindent {\bf (1) From Abiogenesis to the Last Universal Common Ancestor:} There is considerable ambiguity as to when and where life originated, but the environments in which life may have arisen can be broadly classified into ocean- and land-based environments \citep{SAB13,KM18}. The best-known example in the former category is hydrothermal vents \citep{MB08,SHW16},\footnote{Hydrothermal sediments have also been recently posited as the sites of abiogenesis \citep{WHH18}.} while examples in the latter category include geothermal fields \citep{MB12}, beaches \citep{MaL18,AHC18} and intermountain valleys \citep{BKC12}. Another major unknown is the fraction of the surface covered by land and water at the time of abiogenesis. On Earth, the earliest definitive evidence for life is $\sim 3.7$ Ga \citep{LiLo18}, and many recent continental crust models indicate that $\sim 30\%$ of the present-day crustal volume might have existed by this period, as seen from Figures 5 and 6 of \citet{HCD18}. Since we have no way of knowing how continental crust evolved over time on other worlds, we will assume that the fraction $f_w$ did not change significantly since the time of abiogenesis. Thus, if we presuppose an ocean-based origin of life, we find
\begin{equation}
    \mathcal{P}_1 \propto f_w R^2,
\end{equation}
whereas assuming a land-based origin of life leads to
\begin{equation}
    \mathcal{P}_1 \propto f_w\left(1 - f_w\right) R^2.
\end{equation}
\\
Owing to the paucity of observational evidence from fossils and other biomarkers, it is not easy to determine whether the next three steps occurred first on land or in the oceans. However, as these events occurred during the Precambrian period, it seems more plausible that these evolutionary innovations took place in the oceans \citep{Verm17}. The next three megatrajectories are described in the subsequent paragraphs.\\

\noindent {\bf (2) Metabolic Diversification of Bacteria and Archaea:} By $3.4$ Ga, many of the metabolic pathways observed in modern bacteria and archaea had probably evolved. As we have adopted the premise that this step occurred in the oceans, we end up with
\begin{equation}
    \mathcal{P}_{2} \propto f_w R^2.
\end{equation}
\\
\noindent {\bf (3) Evolution  of  the  Eukaryotic Cell:} The origin of eukaryotes has also been debated quite extensively, but the fossil evidence appears to indicate that they had evolved by $1.8$ Ga at the latest \citep{KNo17}. Along the lines of the previous step, the likelihood is given by
\begin{equation}
    \mathcal{P}_{3} \propto f_w R^2.
\end{equation}
\\
\noindent {\bf (4) Complex multicellularity:} Broadly speaking, ``complex multicellularity'' refers to organisms that evolve sophisticated cell differentiation, thereby enabling the efficient transport of oxygen and nutrients among other advantages. Animals, plants and fungi are the chief examples in this category \citep{Kno11}. The likelihood for this step is expressible as
\begin{equation}
    \mathcal{P}_{4} \propto f_w R^2.
\end{equation}
\\
The last two megatrajectories were contingent on the presence of landmasses on Earth, which requires us to adjust the likelihood functions accordingly. A vital point worth reiterating here is that these megatrajectories need not arise in the same order, or in a similar environment, on other worlds. \\

\noindent {\bf (5) Invasion of the Land:} The key thing to note here is that the ``invasion'' refers to the rapid expansion of land plants during the Paleozoic era. The corresponding likelihood is
\begin{equation}
    \mathcal{P}_5 \propto f_w\left(1 - f_w\right) R^2.
\end{equation}
\\
\noindent {\bf (6) Intelligence and Technology:} While intelligence and culture are documented, albeit with some degree of controversy, in cetaceans, the likelihood of advanced technological species emerging in the oceans has been argued to be relatively low \citep{Ste16,LL17}. Assuming that this megatrajectory is land-based, we have
\begin{equation}
    \mathcal{P}_{6} \propto f_w\left(1 - f_w\right) R^2.
\end{equation}
\\
We are now free to define the function $\mathcal{P}_t = \prod_{i=1}^{6} \mathcal{P}_j$ that encapsulates the overall likelihood of technological intelligence emerging on a given world. We are, however, interested in this likelihood relative to the Earth. Hence, we introduce the normalized variable
\begin{equation}
    \Delta_\mathcal{P} = \frac{\mathcal{P}_t}{\mathcal{P}_{t,\oplus}},
\end{equation}
but it must be noted that two different functions exist because the environment associated with abiogenesis is unknown. We use the superscripts `$\ell$' and `$w$' if life originated on land and in the ocean respectively. From the above considerations, we find
\begin{equation}\label{DelPla}
   \Delta_\mathcal{P}^{(\ell)} = \left(\frac{f_w}{f_\oplus}\right)^6 \left(\frac{1- f_w}{1 - f_\oplus}\right)^3 \left(\frac{R}{R_\oplus}\right)^{12},  
\end{equation}
\begin{equation}\label{DelPoc}
   \Delta_\mathcal{P}^{(w)} = \left(\frac{f_w}{f_\oplus}\right)^6 \left(\frac{1- f_w}{1 - f_\oplus}\right)^2 \left(\frac{R}{R_\oplus}\right)^{12}.  
\end{equation}

There are several significant results that can be deduced from (\ref{DelPla}) and (\ref{DelPoc}). One of the primary points worth highlighting is that the likelihood functions are highly sensitive to the size of the world owing to the very strong dependence on $R$ ($\propto R^{12}$). From (\ref{DelPla}), we find that $\Delta_\mathcal{P}^{(\ell)}$ is maximized when $f_w = 2/3$ ($\delta_w = 1/2$) and the value of this quantity becomes $\Delta_\mathcal{P}^{(\ell)} \approx 1.02$. Similarly, by using (\ref{DelPoc}), it is found that the maximum occurs at $f_w = 3/4$ ($\delta_w = 1/3$) corresponding to $\Delta_\mathcal{P}^{(w)} \approx 1.05$. Therefore, our toy model predicts that the Earth's current land-water ratio might be close to optimum insofar the evolution of technological intelligence is concerned. However, worlds larger than the Earth that have the same land-water ratio are potentially superhabitable \citep{HA14} due to the $R^{12}$ dependence alluded to earlier. For example, the recent discovery of Kepler-20b has shown that $R$ can reach a peak value of $\geq 1.87\,R_\oplus$ for rocky planets \citep{BDD16}.

Owing to the greater range spanned by $\delta_w$, we have plotted (\ref{DelPla}) and (\ref{DelPoc}) as a function of this variable in Fig. \ref{FigTLE}. Let us specialize to the case of $R = R_\oplus$ for the sake of simplicity. From the left panel, we see that $\Delta_\mathcal{P}^{(\ell)} > 0.1$ is achieved only when $\delta_w \in \left(0.09,2.17\right)$, or equivalently $f_w \in \left(0.32,0.92\right)$. Using the right hand panel, it can be verified that $\Delta_\mathcal{P}^{(w)} > 0.1$ is fulfilled provided that $\delta_w \in \left(0.04,1.68\right)$, i.e. when $f_w \in \left(0.37,0.96\right)$. Thus, in both these cases, we find that $\delta_w$ spans only about an order of magnitude before the likelihood becomes less than $10\%$ of its maximum value. 

To put it differently, the overall likelihood of technological intelligence becomes $<10\%$ relative to that of the Earth when $f_w < 0.3$. Hence, our model ostensibly implies that the existence of sufficiently wide-ranging oceans is necessary for ensuring the emergence of technological intelligence. This result is a consequence of the fact that all of the critical steps, whether based on the land or in the oceans, depend on the availability of surface liquid water in oceans. Thus, insofar as the extent of land and oceans is concerned, the latter is more crucial than the former. Having said that, the presence of land was crucial for later high-performance evolutionary innovations on Earth \citep{Verm17} and conceivably on other worlds as well \citep{LL17}. Hence, in the limit $f_w \rightarrow 1$, we observe that the functions (\ref{DelPla}) and (\ref{DelPoc}) become vanishingly small. In this respect, our result is different from Sec. \ref{SSecNPPB} because the latter was concerned with non-technological life. 

\section{How common are desert worlds and waterworlds?}\label{SecComm}
A number of theoretical studies have investigated the delivery of water to rocky planets \citep{Liss07,CMP15,RI17}, and the following conclusions appear to be valid: (a) the inventory of water can vary by orders of magnitude \citep{RSM07}, (b) water-rich worlds are quite common \citep{MC15,AB17,Sim17}, and (c) worlds with Earth-like H$_2$O inventories are relatively rare, especially around low-mass stars \citep{TI15,ZDR18}. When it comes to M-dwarfs, determining the water inventories of exoplanets and exomoons is complicated because of the rapid atmospheric escape of H$_2$O \citep{BS17,DHL17}, especially during the long and active pre-main-sequence phase \citep{RK14,LB15}. Moreover, as noted in Sec. \ref{SecIntro}, most worlds that are larger than the Earth are volatile-rich, although the exact cutoff radius is not yet precisely determined.

Despite these recent analyses and discoveries, several ambiguities still exist concerning the probability distribution of water fractions for worlds with varying radii. We define the water fraction as $\mathcal{F}_w = M_w/M$, where $M_w$ is the total mass of water present, not only in the oceans but also in the mantle. For now, we shall focus on worlds with $R \approx R_\oplus$ that have an interior structure similar to that of the Earth. As the probability distribution for $\mathcal{F}_w$ is not fully known, we will consider two representative cases. In both models, we assume that the upper bound on $\mathcal{F}_w$ is $0.5$, and this particular choice is motivated by the fact that some of the TRAPPIST-1 planets may attain such a high value. 

We adopt the relation $M_w = M_{oc} + M_{u}$, where $M_u$ is the mass of water located in the mantle. Clearly, the partitioning of $M_u$ and $M_{oc}$ will depend on the deep-water cycle \citep{Hir06} that, in turn, is governed by geological properties such as temperature, pressure, composition and plate tectonics \citep{SS15,MGR18}. In the case of present-day Earth, we know that $M_{oc,\oplus} \approx 1.4 \times 10^{21}$ kg, but the magnitude of $M_u$ is difficult to estimate. The available evidence appears to indicate that $M_u$ is comparable to, or a few times higher than, $M_{oc}$ \citep{MHY02,Kor08}. We choose the fiducial value $M_u \sim 2 M_{oc,\oplus}$ for the sake of simplicity, and it must be noted that this environment is close to water saturation \citep{FY17}. The next point worth bearing in mind is that the tallest landmass above sea level is Mt. Everest at a height of nearly $9$ km. If we wish to submerge all topographical features above sea level, the mass of water required is $\sim \rho_w \left(4\pi R_\oplus^2\right)(9\,\mathrm{km})$, where $\rho_w$ is the density of water. Adding this mass to the total mass of water already present in Earth's oceans yields a total of $\sim 4.3\,M_{oc,\oplus}$. 

Thus, the end result of the preceding discussion is that waterworlds may form when $M_w > M_u + 4.3 M_{oc,\oplus}$, namely, provided that $M_w > 6.3 M_{oc,\oplus}$. In terms of the water fraction, this inequality amounts to saying that $\mathcal{F}_w > 1.5 \times 10^{-3}$; this result is quite close to the value of $2 \times 10^{-3}$ obtained from a detailed theoretical model \citep{CA14}. A similar analysis can be conducted for desert worlds if we assume that $M_{oc} = 0$ and that $M_u$ attains its saturation value described above. In this case, we find that the condition $\mathcal{F}_w < 4.8 \times 10^{-4}$ may suffice to ensure that the worlds are almost devoid of surface water. For the range $4.8 \times 10^{-4} < \mathcal{F}_w < 1.5 \times 10^{-3}$, it seems plausible that these worlds would consist of both oceans and land. 

The first model that we consider is one where $\mathcal{F}_w$ is uniformly distributed between $0$ and $0.5$.\footnote{Our results remain mostly unchanged if we were to change the lower bound from $0$ to any value that is lower than $4.8 \times 10^{-4}$ by at least one order of magnitude.} From the above data, we arrive at the following three conclusions. The fraction of worlds that lack any water on the surface is $\sim (5 \times 10^{-4})/(0.5) \sim 10^{-3}$. The fraction of worlds that possess a mixture of landmasses and oceans on the surface is $\sim (1.5 \times 10^{-3} - 4.8 \times 10^{-4})/(0.5) \sim 2 \times 10^{-3}$. Finally, the fraction of worlds that have only oceans on the surface is $\sim (0.5 - 1.5 \times 10^{-3})/(0.5) \sim 0.997$. Hence, for this model, the overwhelming majority of worlds ($99.7\%$) are predicted to have no landmasses whatsoever. However, it must be recognized that this outcome arises because the uniform probability distribution is biased in favor of worlds with higher water fractions.

The next model that we consider is a log-uniform distribution, in which $\log \mathcal{F}_w$ is distributed uniformly. The advantage with using such a distribution is that it mitigates the bias associated with the uniform distribution. On the other hand, when using the log-uniform distribution, the lower bound must be non-zero to avoid singular behavior. We adopt the fiducial value of $\sim 5 \times 10^{-6}$ that is roughly two orders of magnitude smaller than the total water fraction of Earth; many numerical simulations tend to use $\sim 10^{-5}$ as the lower bound \citep{RSM07}. We can now repeat the calculations carried out in the previous paragraph. We find that the fraction of worlds devoid of surface water is $\sim 0.4$ whereas the corresponding fraction for worlds without land is $\sim 0.5$. Hence, the fraction of worlds that have both land and oceans on the surface is $\sim 0.1$. 

Thus, we see that the choice of probability distribution for the water fraction has a significant influence in gauging the fraction of worlds with landmasses and oceans on the surface. Based on the right panel of Figure 4 from \citet{MC15}, the fraction of worlds orbiting solar-type stars whose water inventory lies between a few times below and above that of the Earth's content appears to be $\sim 0.2$. Hence, it seems plausible that worlds with Earth-like inventories are uncommon, but are not necessarily very rare. 

\section{Conclusions}\label{SecCon}
In this paper, we have examined how certain properties of extraterrestrial biospheres depend on the fraction of the surface covered by oceans ($f_w$) or land, and on the sizes of these worlds. In the event that the worlds are nearly devoid of surface water, we suggested that the limiting factor could be the availability of liquid water for maintaining complex biospheres. In contrast, if most of the surface is covered by oceans, the availability of nutrients such as phosphates may impose constraints on the total biological productivity, although this factor alone does not rule out the possibility of fairly rich biospheres. By interpolating between these two regimes, we arrived at the following predictions:
\begin{itemize}
    \item The net primary productivity (NPP) and the total biomass vanish in the limit $f_w \rightarrow 0$, but are non-zero (albeit small) when $f_w \rightarrow 1$. The former stems from the fact that surface water is essential for life, but ocean planets could still sustain relatively oligotrophic (nutrient-poor) biospheres with moderate NPP in the latter regime.
    \item The NPP and total biomass do not decrease significantly over a wide range of $\delta_w$ (ratio of land and ocean fractions). The maxima for these quantities occurs at values of $\delta_w$ that are very close to that of the Earth, ostensibly implying that the Earth's topography is close to optimality. It is also found that these two properties have a moderate dependence on the radius $R$ of the world. 
    \item The NPP is connected to the buildup of O$_2$ in the atmosphere when the synthesis of organic compounds occurs via oxygenic photosynthesis. In this case, for Earth-sized worlds, we found that only a narrow range of water fractions ($23$-$88\%$) may, perhaps, enable the buildup of oxygen to levels that are eventually detectable.
    \item The emergence of technological intelligence turned out to be very sensitive to the size of the world, i.e. larger worlds are endowed with a significant advantage. It is also very sensitive to $\delta_w$, and vanishes in the limits of both $f_w \rightarrow 0$ and $f_w \rightarrow 1$. This is because some of the critical evolutionary steps, including the emergence of technological intelligence itself, are presumed to require the presence of land, while others take place in the oceans and thus necessitate liquid water.
    \item For Earth-sized worlds, a water fraction of $30$-$90\%$ might ensure that the likelihood of technological intelligence relative to the Earth is reasonably high. We also found that the Earth's value of $\delta_w$ appears to be close to the optimum with respect to the emergence of technological species, although super-Earths with the same value of $\delta_w$ are probably superhabitable. 
    \item We proposed that worlds with both landmasses and oceans on the surface are rare, although not exceptionally uncommon, in agreement with previous studies.
\end{itemize}
We note that our results are more reliable in the limits $\delta_w \ll 1$ and $\delta_w \gg 1$ relative to the case $\delta_w \sim 1$. For a log-uniform distribution of $\delta_w$, the fraction of worlds with $f_w \in \left(0.3,0.9\right)$ will be $\sim 2.6 \times 10^{-2}$.

Our model effectively includes only two parameters: the water fraction $f_w$ and the size $R$. In actuality, there are a number of abiotic (e.g. ocean pH and photon flux) and biotic (e.g. biomass density and turnover timescales) factors that were held constant. Including the effects of all these variables is a highly challenging endeavor, as they will be coupled to each other through nonlinear feedback mechanisms and are expected to coevolve over time; for instance, biogeochemical cycles ought not be studied in isolation or treated as being constant over time. In our expressions that involved the variable $R$ (planet radius), the majority of the surface area was assumed to be ``habitable'' provided that liquid water exists, but the actual extent of habitable surface area also depends on a number of stellar and planetary factors, for e.g., tidal locking and temperature \citep{Dole,Kas10,SVS17}. 

Furthermore, our analysis did not explicitly incorporate the temporal element, i.e. it was presumed that the worlds have enough time for microbial life and technological intelligence to evolve. However, especially around low-mass stars, factors such as the flux of ultraviolet radiation \citep{RWS17,Manas18,RXT18}, energetic particles associated with stellar flares \citep{SKS07,SWM10,MLin17} and stellar wind-driven atmospheric erosion \citep{DLMC,DJL18,MaLi18} could suppress habitability and effectively lower the duration of time available for biological evolution \citep{TB07,LiLo}. The second aspect regarding temporal evolution is that $f_w$ will change over time owing to the escape of atmospheric water vapor and the transport of liquid water from oceans to the underlying mantle. As a result, the length of time during which the condition $\delta_w \sim 0.1$-$1$ is fulfilled for a given world may have a major influence on its capacity to build up O$_2$ in the atmosphere and evolve technological intelligence. The Earth has been ``lucky'' in this respect, as empirical evidence indicates that this criterion has been satisfied since at least $\sim 3$ Ga \citep{HCD17}.

In spite of the preceding issues, the chief advantage associated with our predictions is that they are readily falsifiable within the next few decades. The ocean and land fractions can, in principle, be determined from photometric observations \citep{FK10,SK18,KNC18}; for example, the presence and extent of oceans is deducible from observations of the ``glint'' effect \citep{WG08,RMC10}. Although there are many potential sources of abiotic O$_2$, methods are being developed to discern whether the atmospheric O$_2$ is biological in nature \citep{MRA18}. Thus, should biogenic O$_2$ be observed mostly on worlds with $f_w \in \left(0.23,0.88\right)$, it may serve as a means of validating our model. The detection of technological intelligence is typically subject to fewer false positives, and there are many strategies that have been identified ranging from electromagnetic signals to megastructures \citep{Tart01,BCD11,WC16,Man18}. Hence, if we find that worlds with distinctive technosignatures are characterized by $f_w \in \left(0.3,0.9\right)$, it could lend some credibility to our model. One crucial point worth recognizing here is that outposts, and therefore technosignatures, can exist even on apparently uninhabitable worlds, especially if the technological species are post-biological in nature \citep{Di03}. 

In summary, there are multiple grounds for contending that the condition $\delta_w \sim 1$ of the Earth is not accidental, since it serves to optimize many facets of its biosphere, including the buildup of O$_2$ in the atmosphere and the emergence of technological intelligence. Interestingly, both of these events were responsible for driving major evolutionary changes on Earth and it is conceivable that they will have a similar effect on other worlds. On account of the fact that worlds with $\delta_w \sim 1$ are relatively uncommon ($\lesssim 1\%$), one might be tempted to argue in favor of the ``Rare Earth'' hypothesis \citep{WaB00}, which favors low numbers of technological species in the Galaxy,\footnote{This would follow from the fact that the fraction of planets that go on to develop technological intelligence in the Drake equation \citep{SS66} can become small.} although this inference is not robust given the current uncertainties. In particular, given that the total number of stars in the Milky Way is very high ($\sim 10^{11}$), even if a very small fraction of stellar systems give rise to complex biospheres with high productivity, the overall numbers for the latter may still end up being relatively large. 

\acknowledgments
We thank our reviewer for the constructive feedback regarding the paper. This work was supported in part by the Breakthrough Prize Foundation, Harvard University's Faculty of Arts and Sciences, and the Institute for Theory and Computation (ITC) at Harvard University.


\end{document}